\documentclass[twocolumn,prl,showpacs,preprintnumbers,amsmath,amssymb,floatfix]{revtex4}
\usepackage{graphicx}
\usepackage{color}

\begin{document}
\title{ Observation of transverse condensation via Hanbury Brown--Twiss correlations}%

\author{Wu RuGway}
\author{ A. G. Manning}
\author{ S. S. Hodgman}
\author{ R. G. Dall}
\affiliation{Research School of Physics and Engineering, Australian National University, Canberra, ACT 0200, Australia}

\author{T. Lamberton}
\author{K. V. Kheruntsyan}
\affiliation{The University of Queensland, School of Mathematics and Physics, Brisbane, QLD 4072, Australia}

\author{A. G. Truscott}
\affiliation{Research School of Physics and Engineering, Australian National University, Canberra, ACT 0200, Australia}

\date{June 6, 2013}%

\begin{abstract}
A fundamental property of a three-dimensional Bose-Einstein condensate (BEC) is long-range coherence, however, in systems of lower dimensionality, not only is the long range coherence destroyed, but additional states of matter are predicted to exist. One such state is a `transverse condensate', first predicted by van Druten and Ketterle [Phys. Rev. Lett. {\bf{79}}, 549 (1997)], in which the gas condenses in the transverse dimensions of a highly anisotropic trap while remaining thermal in the longitudinal dimension. Here we detect the transition from a three-dimensional thermal gas to a gas undergoing transverse condensation by probing Hanbury Brown--Twiss correlations.
\end{abstract}


\maketitle
The behavior of a physical system can be profoundly affected by its dimensionality. A gas of atoms confined in a highly anisotropic harmonic trap, with thermal energy ($k_B T$) that is small compared to the oscillation energies in one or two dimensions, results in the ``freezing out of dynamics" in these dimensions and thus demonstrates strikingly different properties to a three-dimensional (3D) gas \cite{Petrov2004}. For example, in contrast to a 3D gas, effective inter-particle interactions become stronger with decreasing particle number density in the 1D regime enabling a Tonks-Girardeau gas \cite{Girardeau} to be formed. Such a gas corresponds to a system of impenetrable (hard-core) bosons and its behavior in many respects mimics that of free fermions.

The coherence properties of lower dimensional systems are limited primarily due to thermal fluctuations, however, at sufficiently low temperatures quantum fluctuations dominate.  Many nontrivial phases of interest exist for both 1D and 2D ultracold clouds, including the Tonks-Girardeau regime of ``fermionization" in the strongly interacting 1D Bose gas \cite{Kin2004}, quasi-condensation and decoherent quantum regime in weakly interacting Bose gases \cite{Petrov2004,Kher2005,Amer2008,Jacq2011}, transverse condensation in an ideal Bose gas \cite{Druten1997,Armijo2011}, and the Berezinsky-Kosterlitz-Thouless transition in a 2D Bose gas \cite{Krug2007}. 

Transverse condensation occurs in highly anisotropic systems for which the 3D critical temperature ($T_{\rm 3D}^{(0)}$) is higher than that in 1D ($T_{\rm 1D}^{(0)}$). For harmonically trapped gases, these reference temperatures are given, respectively, by $T_{\rm 3D}^{(0)} = \hbar \bar{\omega} [N/g_3(1)]^{\frac{1}{3}} / k_B$, where $g_3(1)\approx 1.202$, and $T_{\rm 1D}^{(0)} = N \hbar \omega_z / [k_B \ln(2 N)] $ \cite{Druten1997}, where $N$ is the total number of atoms and $\bar{\omega} = (\omega_x \omega_y \omega_z)^\frac{1}{3}$ is the geometric mean of the three trap frequencies  $\omega_i$  ($i=x,y,z$). This unusual situation arises when the number of particles is of order the trap aspect ratio, $N \sim \omega_\perp / \omega_z$, where $\omega_\perp \equiv \omega_x =\omega_y$ is the frequency in the transverse direction, assuming a radially symmetric case. As a high-temperature, thermal gas is cooled below the 3D critical transition temperature, the population of excited states in the tightly confining dimension saturates producing transverse condensation of the gas, while the gas remains thermal along the weak trapping dimension. Transverse condensation produces a gas with highly anisotropic coherence properties, in which long-range order is established only in the transverse dimension \cite{Druten1997}. 

The parameter space over which transverse condensation occurs is bounded by the mutual requirements of the gas being in the 3D regime ($k_BT \gtrsim \hbar \omega_\perp$) and for the 1D critical temperature to be less than the 3D critical temperature ($T_{\rm 1D}^{(0)} < T_{\rm 3D}^{(0)}$) \cite{Druten1997}. These mutual requirements can be expressed in terms of two variables, the trap aspect ratio and atom number which can be rewritten in the following form:
\begin{equation}
N[g_3(1)]^{\frac{1}{2}}[\ln(2N)]^{- \frac{3}{2}} < \omega_\perp/\omega_z < N/{g_3(1).}
\end{equation}

The occurrence of the atom number on both extremes of these inequalities leads to a limited parameter space over which transverse condensation can occur; for typical experimentally realizable ratios of $\omega_\perp/\omega_z$ one is restricted to work with a total number of atoms in the hundreds. Moreover, the mutual requirements of the bounds result in transverse condensation occurring in the region approaching the 1D crossover, so that a transversely condensing cloud will always have significant transverse ground state population due to simple 1D physics emerging when the thermal energy $k_BT$ becomes smaller than the transverse excitation energy of $\hbar\omega_\perp$.

The coherence properties of a transversely condensed gas exhibit striking differences from that of a high-temperature thermal gas or a true 3D condensate.  Hence, measuring higher-order correlations along the transverse dimension yields evidence of the transition from a 3D thermal gas to transverse condensation.

\begin{figure}
\begin{center}
\includegraphics[width=3.in,height=3.6in]{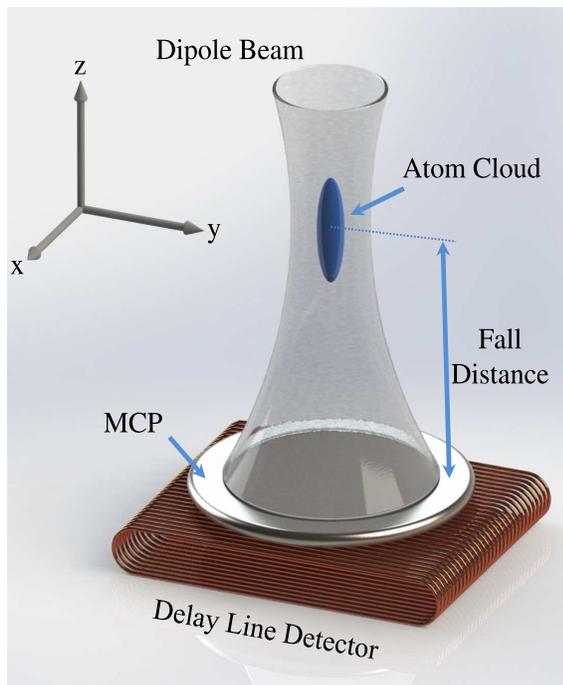}
\caption{Experimental setup used to measure momentum correlations using a multichannel plate (MCP) and delay-line-detector. \label{fig:experiment}}
\end{center}
\end{figure}

Here we report on the observation of transverse condensation in a  cloud of ultracold atoms, first predicted by van Druten and Ketterle in 1997 \cite{Druten1997}, via the measurement of the Hanbury Brown--Twiss correlations \cite{Yasuda,Schel2005,jeltes,Hodgman2011,Guarrera,Manning}. The experimental apparatus is similar to that previously described in \cite{Dall2010, Dall2011} and a simplified schematic of the setup is depicted in Fig.~\ref{fig:experiment}. In detail, $^4$He* atoms are initially evaporatively cooled in our BiQUIC magnetic trap \cite{Dall2007} to just above the 3D Bose-Einstein condensation transition temperature. A dimple is subsequently formed in the trap by overlapping a red detuned focused laser beam with the magnetic trap and ramping up the laser power adiabatically over a period of $100$ ms. This results in the formation of a 
cloud with high phase-space density in the dimple \cite{Stenger1999}.
The magnetic trap is then switched off leaving a small degenerate cloud of atoms ($\sim 10^4$ atoms) in the optical dipole trap at a temperature of $\sim 1 \mu$K. 
To probe different regions of parameter space we reduce the number of atoms in the optical trap using a weak pulse of near resonant light with an intensity of about 100 times lower than the saturation intensity.
As the trap depth is much less than the atomic recoil energy, atoms that absorb a photon exit the trap rapidly with minimal heating of the trapped atoms. The temperature of the remaining trapped thermal atoms was then evaporatively cooled by ramping down the laser power over $200$ ms. 

\begin{figure}
\begin{center}
\includegraphics[width=3.7in,height=3.3in]{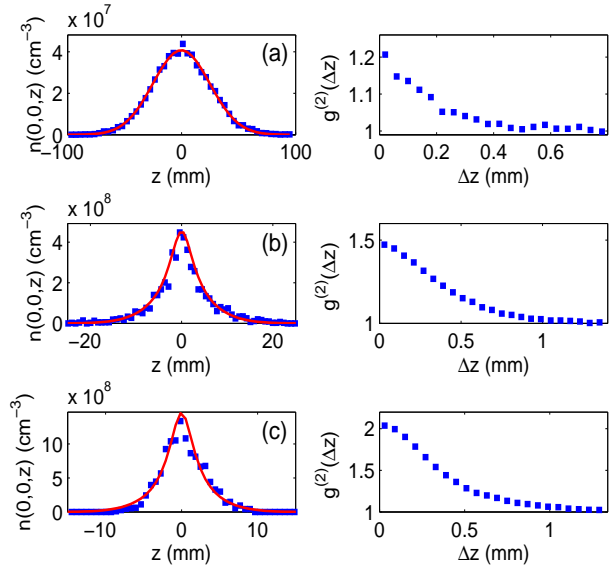}
\caption{(Left Panels) Longitudinal slices, $n(0,0,z)$, of the density distribution of the cloud at the detector
and theoretical fits (solid red lines, ideal Bose gas model) for the three sets of data: (a)  $N=2800$, $T/T_{\rm 3D}^{(0)}=2.74$; (b) $N=820$, $T/T_{\rm 3D}^{(0)}=0.72$; and (c)  $N=370$, $T/T_{\rm 3D}^{(0)}=0.52$. (Right panels) The corresponding longitudinal second order-correlation functions $g^{(2)}(\Delta z)$. To determine the atom number in each cloud we use the quantum efficiency (QE) as a fitting parameter allowing it to vary in the range 0.2 to 0.3, which is consistent with QE measurements we have made for our detector. 
} 
\label{fig:tof_gt}
\end{center}
\end{figure}

\begin{figure}[tbh]
\begin{center}
\includegraphics[width=3.6in,height=3.1in]{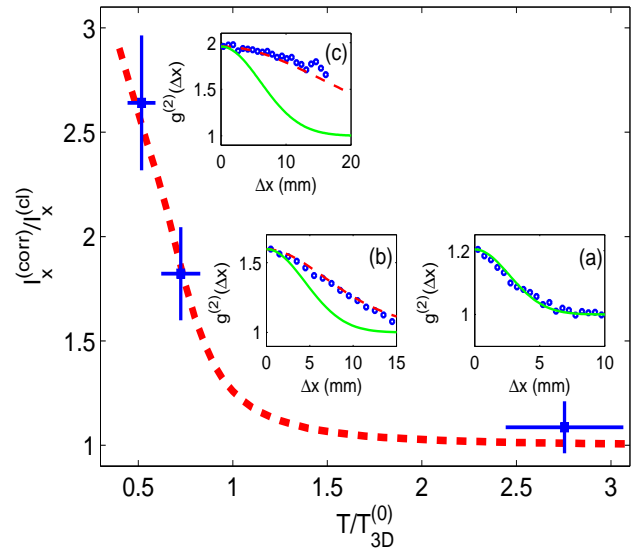}
\caption{Scaled transverse correlation length $l_x^{(\mathrm{corr})}/l_x^{(\mathrm{cl})}$ as a function of $T/T_{\rm 3D}^{(0)}$. Three points are shown for three different parameters regimes, described in the text as (a), (b), and (c). The insets show the measured second-order correlation functions $g^{(2)}(\Delta x)$ for each case. The correlation length $l_x^{(\mathrm{corr})}$ is defined as the rms width of the function $g^{(2)}(\Delta x)-1$ and is scaled with respect to a reference length scale -- the correlation length at the detector for a classical (Boltzmann) ideal gas given by the rms width of a Gaussian, $l_x^{(\mathrm{cl})} = \hbar \omega_x t_{\mathrm{ToF}} /\sqrt{2m k_B T}$ \cite{Gomes2006}. The dashed (red) and solid (green) lines in the insets are theoretical curves from the ideal Bose gas and the classical ideal gas models, respectively,  produced using the number and temperature values extracted from fits to the data, in (a) these two lines overlay each other. The dashed (red) curve in the main figure is produced using ideal Bose gas theory, however, the theoretical curve below $T_{\rm 3D}^{(0)}$ has a weak dependence on atom number which is not constant for (b) and (c), thus we show  the average curve which results from using the parameters of (b) and (c). To calculate the rms width of the measured correlation function, we use an ideal Bose gas best fit.  The error bars on the data points in the main graph represent our uncertainty in $T$ (10\%), $N$ (25\%), $\omega_{i}$ (10\%), and the measured correlation length (5\%). 
}
\label{fig:trans_corr}
\end{center}
\end{figure}

The optical dipole trap is highly anisotropic and aligned with its weak axis in the direction of gravity ($z$-axis). To enter the transverse condensation regime, we require that the atom number be of order the aspect ratio of the trap. Our datasets span three distinct regimes shown in Figs.~\ref{fig:tof_gt} and \ref{fig:trans_corr} below (to be discussed shortly): (a) a 3D thermal gas with $N\!=\!2800$, $T\!=\!1.7$~$\mu$K, and $T/T_{\rm 3D}^{(0)}\!=\!2.74$; (b) a partially transversely condensed cloud with $N\!=\!820$, $T\!=\!155$~nK, and $T/T_{\rm 3D}^{(0)}\!=\!0.72$; and (c) a more pure transversely condensed cloud with $N\!=\!370$, $T\!=\!63 $nK, and $T/T_{\rm 3D}^{(0)}\!=\!0.52$.
For the datasets (a), (b), and (c), the trapping frequencies are $(\omega_x,\omega_y,\omega_z)/2\pi\!=\!(5400,4500,38)$ Hz, $(2700,2300,21)$ Hz, and $(2350,1700,13)$ Hz, respectively. Each dataset represents of order $1000$ realisations of the experiment.

Momentum correlation functions are measured in the far field by switching off the optical trap, thus allowing the atoms to fall for $t_{\mathrm{ToF}}=416$ ms onto a multichannel plate and delay-line-detector located $848$ mm below where the arrival time and position of each atom is measured. Single $^4$He* atoms are efficiently detected due to their large internal energy $19.8$~eV and long lifetime $7860$~s \cite{Hodgman2009}. The detector has a spatial resolution of $\sim 100$ $\mu$m in the $(x,y)$-plane and an effective spatial resolution in the vertical direction ($z$-axis) of the order of nanometres (temporally a few nanoseconds). Using the resulting single atom arrival positions and times an average many-body correlation function can be calculated for both the transverse directions ($x$ or $y$-axis) and for the longitudinal direction ($z$-axis). 

The second-order correlation function along one of the transverse directions (here chosen to be the $x$-direction) were calculated using atom counting bins $6$mm $\times$ $160$ $\mu$m in size for the ($y,z$)-plane and using the following equation:
\begin{equation}
g^{(2)}(\Delta x) = \frac{\int d^3\mathbf{r}G^{(2)}(\mathbf{r},\mathbf{r}+\mathbf{e}_x\Delta x)}{\int d^3\mathbf{r} \langle n(\mathbf{r}) \rangle \langle n(\mathbf{r}+\mathbf{e}_x \Delta x)\rangle}.
\end{equation}
Here, $\mathbf{e}_x$ is a unit vector in the $x$-direction, $G^{(2)}(\mathbf{r}_1,\mathbf{r}_2)=\langle n(\mathbf{r}_1)n(\mathbf{r}_2) \rangle $ is the unnormalised two-body correlation function, and $n(\mathbf{r})$ is the spatial density distribution on the detector. The longitudinal correlation functions $g^{(2)}(\Delta z)$ were calculated by binning individual atom detection events, in the horizontal  ($x,y$)-plane, into $5$mm $\times$ $5$mm bins for (a), and $1$cm $\times$ $1$cm bins for (b) and (c). To calculate $g^{(2)}(\Delta z)$ we use a similar equation to the one above, simply replacing $\Delta x$ with $\Delta z$ and $\mathbf{e}_x$ with $\mathbf{e}_z$.

The measured atomic density profiles, along the z-axis, at the detector and the corresponding longitudinal correlation functions for each datasets are shown in Fig. \ref{fig:tof_gt}. The density profiles can be transformed to the  momentum distributions of the trapped sample by applying the ballistic expansion transformation $\hbar k_z=mz/t_{\mathrm{ToF}}$.
Note that, as the temperature of the gas is cooled below $T_{\rm 3D}^{(0)}$, there is no evidence of a true Bose-Einstein condensation; the shapes of the time-of-flight profiles in all dimensions remain thermal and are well fitted by the ideal Bose gas model in a 3D harmonic trap (solid red lines) \cite{Naraschewski1999,Gomes2006}. These fits were also used to derive the quoted temperature of each sample. We will return to the discussion of the longitudinal correlation functions (right panels in Fig.~\ref{fig:tof_gt}) after discussing the correlation measurements in the transverse direction.

The transverse condensation transition is illustrated in Fig.~\ref{fig:trans_corr}, where we plot a scaled transverse correlation length  for three different degeneracy parameters. In the high-temperature 3D regime (a), the second-order correlation function $g^{(2)}(\Delta x)$ is simply a Gaussian consistent with a classical Boltzmann gas model \cite{Gomes2006}. As the gas is cooled further, we enter the transverse condensation regime (b) and this is reflected in the difference between the dashed (ideal 3D Bose gas model) and solid (classical Boltzmann gas) curves. In this case, an ideal Bose gas model yields a transverse ground-state population of 51\%, however applying a simple Boltzmann calculation results in a population of 29\%.
Finally, in (c) after further cooling, we observe a dramatic increase of the transverse correlation length, which now exceeds the cloud size at the detector, and we calculate a much higher (82\%) transverse ground-state population from the ideal Bose gas model (as opposed to 61\% from the Boltzmann gas). As we mentioned earlier, the temperatures were derived from a fit to the longitudinal time-of-flight density profile, and importantly no discernible longitudinal ground state fractions were observed even at temperatures as low as $\sim \!0.5 T_{\rm 3D}^{(0)}$ -- in agreement with the prediction for a transversely condensed gas \cite{Druten1997}.

The critical transition temperature $T_{\rm 3D}^{(0)}$, which we use as a reference temperature, corresponds to the 3D condensation temperature in the thermodynamic limit \cite{Druten1997}. Although there is no unique definition of the transition temperature for finite-size systems the effect of having a finite number of atoms is known to result in lowering the critical temperature to \cite{Ketterle1996}
\begin{equation}
T^*_{3D} = T_{\rm 3D}^{(0)} \left(1-\frac{0.7275 \sum _i \omega_i}{3N^{\frac{1}{3}}(\prod _i \omega_i)^{\frac{1}{3}}}\right).
\end{equation}
This explains why the dramatic increase of the transverse correlation length in Fig. \ref{fig:trans_corr} occurs at temperatures somewhat lower than $T_{\rm 3D}^{(0)}$. We also point out that the finite-size effects for our clouds of relatively small total number of atoms render the transition to the transverse condensate as a smooth crossover rather than as a sharp phase transition.

The realisation of a nearly pure transverse condensate, as in Fig. \ref{fig:trans_corr} (c), enables us to produce matter waves with somewhat unique correlation properties. This is due to the gas being transversely coherent, but longitudinally incoherent, where interference between the many occupied longitudinal modes occurs over the entire transverse extent of the cloud. Indeed, these modal properties led to transverse condensation to be alternatively described as `multimode condensation' \cite{Druten1997}.
The peak amplitude for the coldest cloud (c) is in good agreement with the maximum value of $2!$ \cite{Dall2013} expected for the second-order correlation function for a thermal gas, while as the temperature increases and the correlation length shortens in (b) and (a), the finite bin sizes used in our data analysis result in a reduction of the peak bunching amplitude \cite{Schel2005,jeltes,Hodgman2011,Guarrera,Manning}. Note, that differences between the peak bunching amplitude observed transversely and longitudinally can result from different bin choices in those axes. The respective longitudinal correlation lengths are far smaller than the cloud size. This absence of long-range order longitudinally demonstrates that the gas remains thermal in the $z$-dimension while condensing transversely.

We emphasise that the nature of the critical transition to a transversely condensed gas, which enabled us to measure the maximum second-order correlation amplitude of $g^{(2)}(0)=2$, is a quantum degeneracy driven (rather than interaction driven \cite{Bouchoule2007}) transition in an ideal Bose gas confined to a highly anisotropic trap. As discussed in Refs. \cite{Druten1997} and \cite{Armijo2011}, this occurs due to the saturation of population in the transversely excited states hence the term `transverse condensation'. Our measurements of the transverse properties of the gas, such as the fraction of the atoms in the transverse ground state and the transverse correlation function, are indeed in excellent agreement with the predictions of the theory of a harmonically trapped ideal Bose gas. On the other hand, the measured shape of the longitudinal second-order correlation function (see Fig.~\ref{fig:tof_gt}), especially well below the critical temperature, appears to be intermediate between the theory of a highly-degenerate ideal Bose gas and a weakly interacting 1D quasicondensate \cite{Bouchoule2012}.

For example, for the coldest sample [Fig.~2(c)], the measured longitudinal correlation length on the detector ($\sim 0.4$ mm) is about 4 times larger than the prediction of the ideal Bose gas theory. The origin of this discrepancy is not well understood, however, given our small clouds, we expect the finite-size effects to be significant which means that the physics in the longitudinal dimension is dominated by broad crossovers between that of a pure noninteracting gas and a weakly interacting gas approaching the quasicondensate regime.

As a crude alternative estimate of the longitudinal momentum-momentum correlation length we can use the theory of a uniform 1D Bose in the quasicondensate regime \cite{Bouchoule2012}. Applying it to the central part of our trapped cloud, assuming a local density approximation, one can estimate the characteristic momentum correlation length to be given by $\Delta k_z \!\sim\! (2l_\phi)^{-1}$ (in wave-number units), where $l_\phi =\hbar^2 n_{\rm 1D}/mk_BT$ is the characteristic  phase coherence length and $n_{\rm 1D}$ is the local 1D density equal to $n_{\rm 1D}\simeq 2.0$ atoms/$\mu$ in our case (c). Converting this to the spatial correlation length after time-of-flight expansion gives a longitudinal correlation length on the detector of $0.85$ mm. Thus our measured value of $\sim \!0.4$ mm is indeed intermediate between the predictions of the ideal Bose gas theory and the theory of a 1D quasicondensate. Our estimates of the 1D dimensionless interaction parameter $\gamma=mg/\hbar^2n_{\rm 1D}$ (where $g=2\hbar\omega_\perp a$ is the 1D coupling and $a$ is the 3D $s$-wave scattering length) and the dimensionless temperature $\tau=2mk_BT/\hbar^2n_{\rm 1D}^2$ \cite{Kher2005} also support this conclusion. Indeed, the estimated values are $\gamma\simeq 0.006$ and $\tau\simeq 0.26$ for our case (c). The temperature $\tau$ in this case is just three times larger than the crossover boundary ($\tau\sim \sqrt{\gamma}$) separating the so-called decoherent quantum (or a nearly ideal Bose gas) regime ($\tau \!\gg \!\sqrt{\gamma}$) and the quasicondensate regime ($\tau\!\ll \!\sqrt{\gamma}$) \cite{Kher2005,Deuar2009}.

In conclusion, we have used Hanbury Brown--Twiss correlations to demonstrate the existence of transverse condensation, a state of matter that has coherence properties of a condensate along its transverse dimensions, while exhibiting thermal characteristics along its remaining longitudinal dimension. These coherence properites make a transversly condensed Bose gas an ideal candidate for high resolution atomic coherence tomography, analogous to optical coherence tomography \cite{Huang1991}. Moreover, the precise measurements of second-order correlation lengths are a useful tool for discerning phase diagrams of various ultracold atom systems, especially systems of lower dimensionality where the physics is dominated by the stronger role of quantum fluctuations and correlations, leading to a number of nontrivial phases absent in 3D systems.

A.G.T and K.V.K. acknowledge the support of the Australian Research Council through the Future Fellowship grants FT100100468 and FT100100285.

\end{document}